\documentclass[preprint]{vgtc}               % preprint
\graphicspath{{figures/}{pictures/}{images/}{./}} % where to search for the images

\usepackage{times}                     % we use Times as the main font
\usepackage{tabu}                      % only used for the table example
\usepackage{booktabs}                  % only used for the table example
\usepackage{lipsum}                    % used to generate placeholder text
\usepackage{mwe}                       % used to generate placeholder figures

\usepackage{mathptmx}                  % use matching math font
\usepackage{graphicx}
\usepackage{subfigure}
\usepackage{amsfonts,amsmath}
\usepackage{multirow}
\usepackage{svg}

\onlineid{2299}

\vgtccategory{Research}

\vgtcinsertpkg

\title{TimeGazer: Temporal Modeling of Predictive Gaze Stabilization for AR Interaction}

\author{
  Yaozheng Xia$^{1}$,
  Zaiping Zhu$^{2}$,
  Bo Pang$^{3}$,
  Sheng Li$^{3}$,
  Shaorong Wang$^{1}$ \\
  \\
  $^{1}$Beijing Forestry University, Beijing, China \\
  $^{2}$Bournemouth University, National Centre for Computer Animation, Poole, Bournemouth, United Kingdom \\
  $^{3}$Peking University, Beijing, China
}

\teaser{
  \centering
  \includegraphics[width=0.85\linewidth]{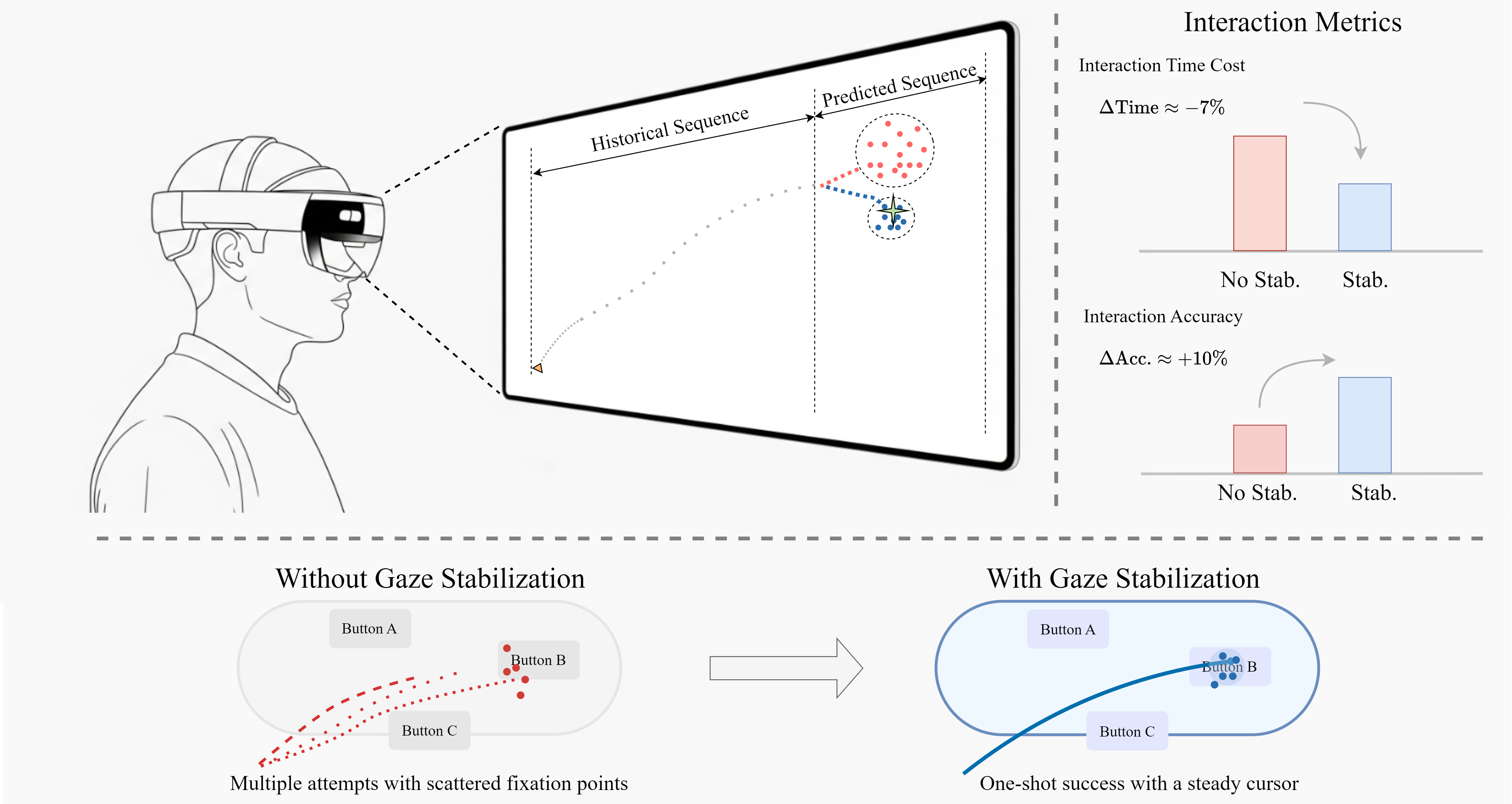}
  \caption{A user wearing an AR HMD has their gaze tracked from discovery (yellow triangle) to target (green star). TimeGazer leverages historical gaze to predict a more precise future path (blue) that stabilizes raw gaze data (red), yielding more accurate fixations, higher hit rates, and reduced task time (see \cref{tab:user_study}.)}
  \label{fig:teaser}
}

\abstract{
Gaze stabilization is critical for enabling fluid, accurate, and efficient interaction in immersive augmented reality (AR) environments, particularly during task-oriented visual behaviors. However, fixation sequences captured in active gaze tasks often exhibit irregular dispersion and systematic deviations from target locations, a variability primarily caused by the combined effects of human oculomotor physiology, insufficient AR headset tracking and calibration accuracy, and environmental disturbances, undermining interaction performance and visual engagement.
To address this issue, we propose TimeGazer, which reformulates gaze stabilization as a sequence-to-sequence temporal regression problem, predicting idealized fixation trajectories for the target-fixation phase from historical gaze dynamics in the search phase. We present a synthetic data generation and blending strategy that produces spatially concentrated, target-centered fixation references aligned with task objectives, substantially enriching the training space and enhancing model generalization.
We train and evaluate TimeGazer on a hybrid dataset of real and augmented gaze sequences collected via Microsoft HoloLens 2 from 54 participants across multiple prediction horizons. Through the user study, statistical results demonstrate that TimeGazer significantly improves interaction accuracy and reduces completion time, confirming that temporal modeling of predictive gaze stabilization can strengthen attentional consistency and responsiveness in task-driven AR interaction. These findings highlight the broader potential of TimeGazer for advancing adaptive gaze-based interfaces and temporal modeling research in immersive systems.

} % end of abstract


\begin{document}


\firstsection{Introduction}

\maketitle
Eye-tracking technology has become a pivotal component of augmented reality (AR) systems, enabling intuitive human–computer interaction through gaze-based control and attention analysis \cite{krafka2016eye}. In AR applications such as gaze typing and augmented reading, the fixation phase is critical for accurately inferring user intent and enhancing interaction efficiency\cite{Saran_Alber_Zhang_Paradiso_Bragg_Langford_2025}. However, human gaze behavior alternates between saccades—rapid eye movements that shift fixation—and fixations, where the eyes remain relatively stable to acquire visual information\cite{OPTICAN200925}. Despite this stability, raw gaze data collected during fixation often exhibits considerable dispersion and deviation from the intended target, arising from human physiological factors, hardware limitations, and environmental conditions\cite{bach2012control,Marquardt_Steininger_Trepkowski_Weier_Kruijff_2024}. Such dispersion necessitates longer fixation times to compensate for reduced spatial precision, thereby increasing cognitive load and visual fatigue, which ultimately degrades user experience\cite{wang2019assessment,liu2022assessing}. More specifically, unintended gaze shifts from saccades to fixations (or occurring during fixations) can severely impair the user experience and compromise interaction performance.

Since gaze naturally forms a spatiotemporal sequence\cite{Burlingham_Sendhilnathan_Wu_Murdison_Proulx_2024}, its velocity, acceleration, and dispersion patterns—timestamped and correlated across points—carry rich cues about user intent and perceptual stability\cite{Jindal_2024_CVPR}. This has motivated extensive research on scanpath prediction, which aims to model and anticipate sequential gaze behaviors\cite{Qiu_Rong_Liang_Tu_2023, Wang_Bâce_Bulling_2024, Chen_Jiang_Zhao_2024, Yang_Mondal_Ahn_Xue_Zelinsky_Hoai_Samaras}. However, most of these studies primarily address scanpaths in free-viewing scenarios within static visual scenes, rather than gaze sequences driven by actively controlled tasks. In contrast, goal-directed or volitional gaze tasks such as reading\cite{deng2023eyettention, kaakinen2010task, reichle2010eye} or mid-air typing\cite{Saran_Alber_Zhang_Paradiso_Bragg_Langford_2025,Hu_Dudley_Kristensson_2024, Zhao_Pierce_Tan_Zhang_Wang_Jonker_Benko_Gupta_2023} are inherently constrained by directional and spatial patterns of eye movements\cite{boisvert2016predicting,henderson2013predicting} (e.g., line-by-line text reading or navigating a virtual keyboard), limiting their generality for broader gaze-based interaction. To alleviate this gap, we aim to optimize the gaze experience under weak directional constraints in active attention tasks, enabling so-called gaze stabilization to facilitate fundamental human–computer interaction events such as object selection in VR/AR environments.

The sequential nature of gaze data naturally aligns with time series analysis, which has proven highly effective in capturing temporal dependencies across domains such as speech recognition\cite{Temporal_Modeling_Matters} and financial forecasting\cite{wang2021hierarchical}. By leveraging the temporal dependencies inherent in gaze data, numerous approaches have been proposed for modeling eye movement sequences. Scanpath prediction methods have demonstrated high accuracy, but they predominantly focus on free-viewing scenarios rather than task-driven settings\cite{Wang_Zhang_Dodgson_2024, Zhu_Zhang_Min_Zhai_Yang_2025}. Similarly, gaze estimation and prediction methods achieve impressive performance, yet most operate from a third-person perspective and lack consideration of egocentric environments\cite{gupta2024mtgs,Jindal_2024_CVPR}. Recent studies have begun to address active attention tasks in egocentric scenarios\cite{Liu_Cheng_Sun_Wu_Song_Sun_Zhang_2025,lai2024listen}; however, these works often rely on multimodal frameworks, which not only obscure the primary role of visual attention in such tasks but also introduce practical challenges for deployment on AR/VR devices.

To address this gap, we introduce TimesGazer, a novel approach explicitly exploits the rich information embedded in gaze trajectories over time by leveraging temporal sequence modeling. The spatiotemporal dynamics of gaze—particularly velocity changes and dispersion trends—encode user intent and attentional stability, which our model captures to predict optimized fixation points and improve AR interaction reliability.
To the best of our knowledge, this is the first to explicitly harness temporal dynamics for refining gaze positions during fixation. Our approach achieves robust gaze stabilization through predictive modeling to ensure both accuracy and efficiency in active attention tasks in AR system. Our primary objective is to reduce gaze dispersion and minimize the offset from target points, thereby enhancing the reliability of gaze-based interactions in AR applications. To achieve this, we develop a sequence-to-sequence model that exploits historical gaze trajectories from preceding saccadic phases to predict optimized fixation points in the subsequent fixation phase of active tasks. For training, we collected eye movement data from diverse participants performing controlled fixation tasks under minimally distracting conditions, constructing paired datasets for supervised learning. Our model is deployed and validated on AR mobile devices with significant performance improvement.

The main contributions of this work are as follows:
\begin{itemize}
   
\item We propose a predictive gaze stabilization framework for AR that reformulates stabilization as a sequence-to-sequence temporal regression problem, refining fixation points without relying on scene semantics or multimodal cues, and generalizing across individuals by capturing shared temporal patterns in gaze behavior.

\item We present a synthetic data generation and blending strategy that creates target-centered augmented fixation sequences, expanding training diversity and improving model generalization for temporal gaze dynamics.

\item We curate a purpose-built temporal gaze dataset of paired saccade–fixation sequences collected from diverse participants under controlled conditions, providing a robust foundation for supervised learning in active attention tasks.

\item We optimize and deploy TimeGazer on mobile AR headsets, achieving low-latency, real-time inference and demonstrating significant gains in interaction accuracy and task completion time across multiple prediction horizons.
\end{itemize}

%1.\textbf{A novel framework for gaze stabilization in AR} – We introduce TimesGazer, the first temporal modeling approach designed to refine fixation-phase gaze points without relying on image semantics or multimodal attention cues. This framework effectively reduces gaze dispersion and improves spatial alignment with target objects, enabling reliable gaze-based interaction in AR environments.

%2.\textbf{A purpose-built eye-tracking dataset} – We collect and curate a high-quality dataset of eye movements from diverse participants performing controlled fixation tasks under minimally distracting conditions. This dataset ensures broad applicability across user groups and provides a solid foundation for supervised learning of temporal gaze dynamics.

%3.\textbf{Deployment and validation on AR hardware} – We optimize and deploy the trained model on mobile AR headsets (e.g., Microsoft HoloLens 2), achieving real-time inference and demonstrating its effectiveness in interactive AR scenarios.

\section{Related Works}
First of all, methods of gaze tracking are presented. Then, the temporal analysis model and its applications are summarized.
\subsection{Gaze Tracking}
Gaze tracking in AR systems has evolved from hardware-centric methods like Pupil Center Corneal Reflection (PCCR)\cite{ohno2002freegaze} to modern deep learning approaches with temporal modelling capabilities. PCCR remains the standard in commercial XR headsets such as Microsoft HoloLens 2, HTC Vive Pro Eye, and Meta Quest Pro for its robustness and slippage compensation, though it increases hardware cost and calibration complexity. Recent appearance-based approaches using CNNs, LSTMs, and Transformers treat gaze as a spatiotemporal signal, enabling learning-based estimation across frames\cite{Dan2010,liu2022eye}. While recent architectures—such as ViT-based and hybrid CNN–Transformer models—have demonstrated strong performance in gaze estimation \cite{Liu_Duinkharjav_Sun_Zhang_2025, Qin_Zhang_Sugano_2025}, studies in VR contexts consistently show that gaze dispersion during the fixation phase degrades interaction precision, elevates cognitive load, and contributes to visual fatigue \cite{Nasri_Kosa_Chukoskie_Moghaddam_Harteveld_2024}.

\subsection{Temporal Analysis Model}
Temporal modeling has evolved from early statistical methods to advanced deep learning approaches, each addressing different aspects of sequential dependency. Classical techniques such as ARIMA\cite{Ke_Meng_Finley_Wang_Chen_Ma_Ye_Liu_2017} and Holt-Winters\cite{Chatfield_1978} assume fixed temporal patterns and struggle with highly dynamic or nonlinear behaviors. To overcome these limitations, machine learning models like LightGBM\cite{Ke_Meng_Finley_Wang_Chen_Ma_Ye_Liu_2017} and XGBoost\cite{Chen_Guestrin_2016} introduced greater flexibility, though they remain limited in capturing complex long-range dependencies. Deep learning models further advanced temporal analysis through diverse architectures: MLP-based models encode temporal dependencies into fixed parameters, offering computational efficiency but limited capacity for long-term relationships\cite{oreshkin2019n,challu2023nhits,zeng2023transformers,zhang2023MLPST}. Temporal Convolutional Networks (TCNs) utilize dilated causal convolutions for parallelized modeling of local and mid-range patterns, though receptive fields require careful tuning\cite{bai2018empirical,sen2019think,liu2022scinet}. RNN and LSTM architectures capture sequential dynamics through recurrent state transitions and support variable-length sequences, but suffer from limited parallelism and difficulty modeling very long dependencies\cite{lai2018modeling,zhao2020rnn,rusch2021unicornn}. Transformers introduce self-attention for global dependency modeling and multi-scale temporal adaptation, albeit with substantial computational overhead\cite{wu2021autoformer,zhou2022fedformer}. For instance, PatchTST\cite{nie2022time} leverages sub-sequence patching to enhance long-sequence modeling, while iTransformer\cite{liu2023itransformer} treats each variable’s history as an attention token, improving inter-variable dependency capture and long-term context understanding.

These advancements highlight the strength of temporal analysis in handling complex sequential patterns, making it a compelling foundation for gaze-related research in AR environments.

\subsection{Applications of Temporal Models in Gaze Tracking}
Temporal models are widely applied in gaze-related tasks. In gaze estimation, spatio-temporal models exploit eye, head, and body dynamics for accurate current fixation, but rarely predict future gaze \cite{Jindal_2024_CVPR,Nonaka_2022_CVPR}. In gaze prediction, models leverage historical eye movements, scene context, and user behavior to forecast future gaze targets or trajectories. For example, Hu et al. \cite{hu2021fixationnet} predict users’ fixation points in task-driven VR settings, demonstrating the feasibility of gaze forecasting for interactive virtual environments. Gupta et al.\cite{gupta2024mtgs} proposed MTGS, a transformer-based model for multi-person gaze prediction in third-person scenes, but not for egocentric views. Recent egocentric approaches, such as Lai et al.\cite{lai2024listen} and Liu et al.\cite{Liu_Cheng_Sun_Wu_Song_Sun_Zhang_2025}, incorporate audiovisual cues to predict gaze trajectories; however, their multi-modal pipelines are often too heavy for real-time AR devices. In scanpath prediction, methods like ScanDMM, ScanTD, and ScanDTM \cite{Sui_Fang_Zhu_Wang_Wang_2023, Zhu_Zhang_Min_Zhai_Yang_2025,Wang_Zhang_Dodgson_2024} model temporal dependencies between discrete fixations, yet mostly under free-viewing or weakly constrained attention, with little focus on goal-directed or active fixation tasks. This gap motivates our work on stabilizing active fixations in AR via temporal modeling and predicting.

\section{Data Collection}
\subsection{Stimuli}
\begin{figure}[htbp]
	\centering
	\subfigure {\includegraphics[width=.22\textwidth]{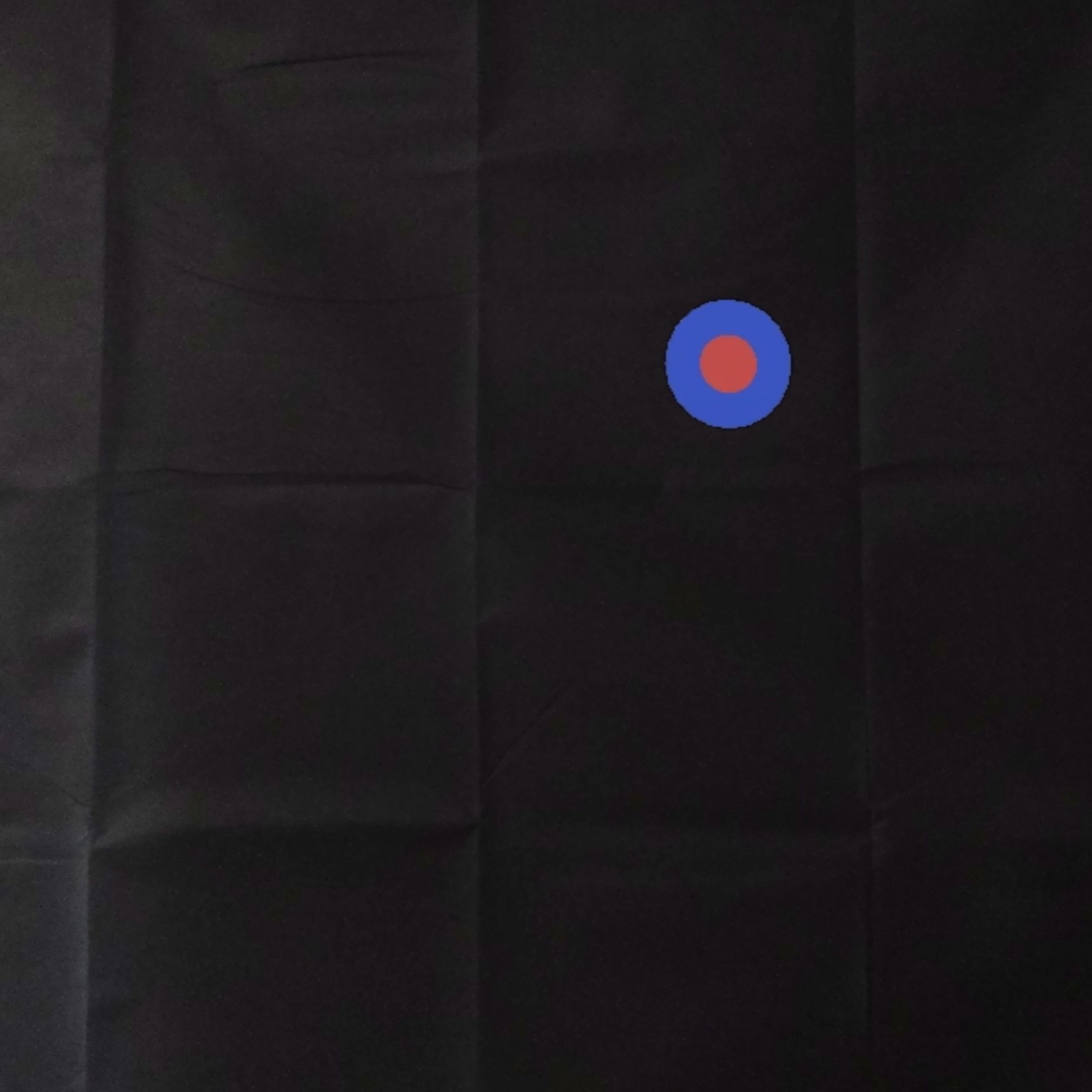}}
	\subfigure {\includegraphics[width=.22\textwidth]{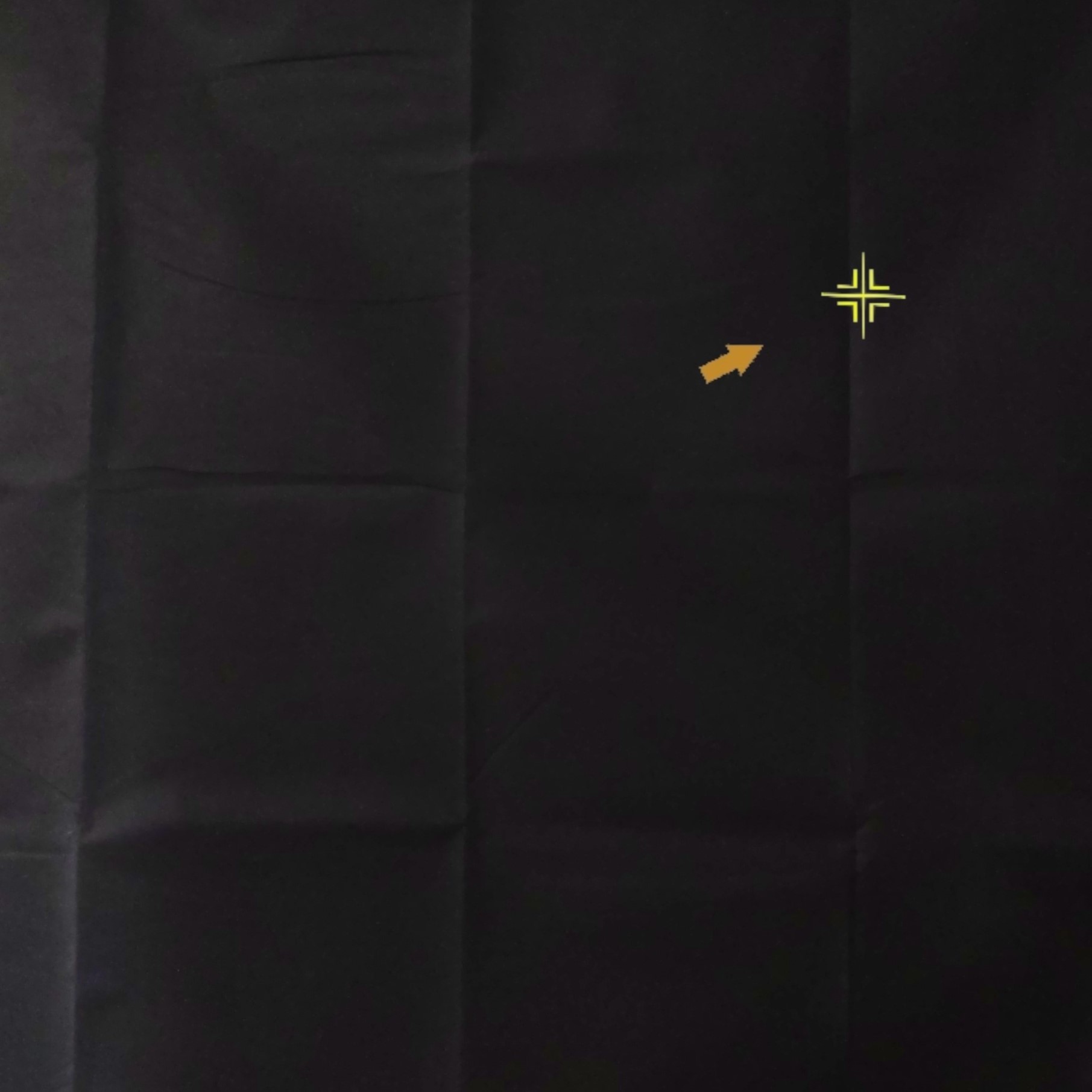}}
	\caption{Real-world background and visual elements in the visual environments.}
	\label{fig_2}
\end{figure}

To collect gaze data, we constructed an AR environment as the experimental stimulus (see  \cref{fig_2}). The real-world background, positioned 3 meters away and consisting of a large, plain black mat sufficient to encompass the entire virtual scene, was paired with a virtual scene also positioned 3 meters in front of the headset. This virtual scene measured 2 × 2 meters and was deliberately kept minimal to reduce potential confounding factors. Only three types of visual elements were presented: (1) a red-blue concentric circle (radius: 0.1 m) serving as the start marker for each trial, (2) a tan arrow indicating the participant's intended gaze shift direction, and (3) a yellow cross (arm length: 0.1 m) serving as the fixation target, at whose center participants were instructed to maintain steady fixation. This minimalist design reduces scene complexity, thereby enhancing the discriminability of fixation events and enabling a clearer separation of physiological eye movements (e.g., saccades, microsaccades) from task-driven gaze behavior \cite{holmqvist2011eye,salvucci2000identifying}.

\subsection{Apparatus and Participants}
Our data collection was conducted on a PC equipped with an AMD Ryzen™ 5 7500F CPU (3.70 GHz) and an NVIDIA GeForce RTX 2080 GPU. The experimental setup was implemented on a Microsoft HoloLens 2 headset, which integrates an optical eye-tracking system operating at 60 Hz with an accuracy of approximately 0.5°. Head motion was simultaneously recorded using the HoloLens 2’s built-in head tracking system at a sampling rate of 60 Hz. The experimental scenes were rendered in real time using the Unity3D engine, and custom Unity scripts were developed to log task-related gaze and head information at 60 Hz. 
 
\begin{figure}[htbp]
	\centering
	\subfigure {\includegraphics[width=.22\textwidth]{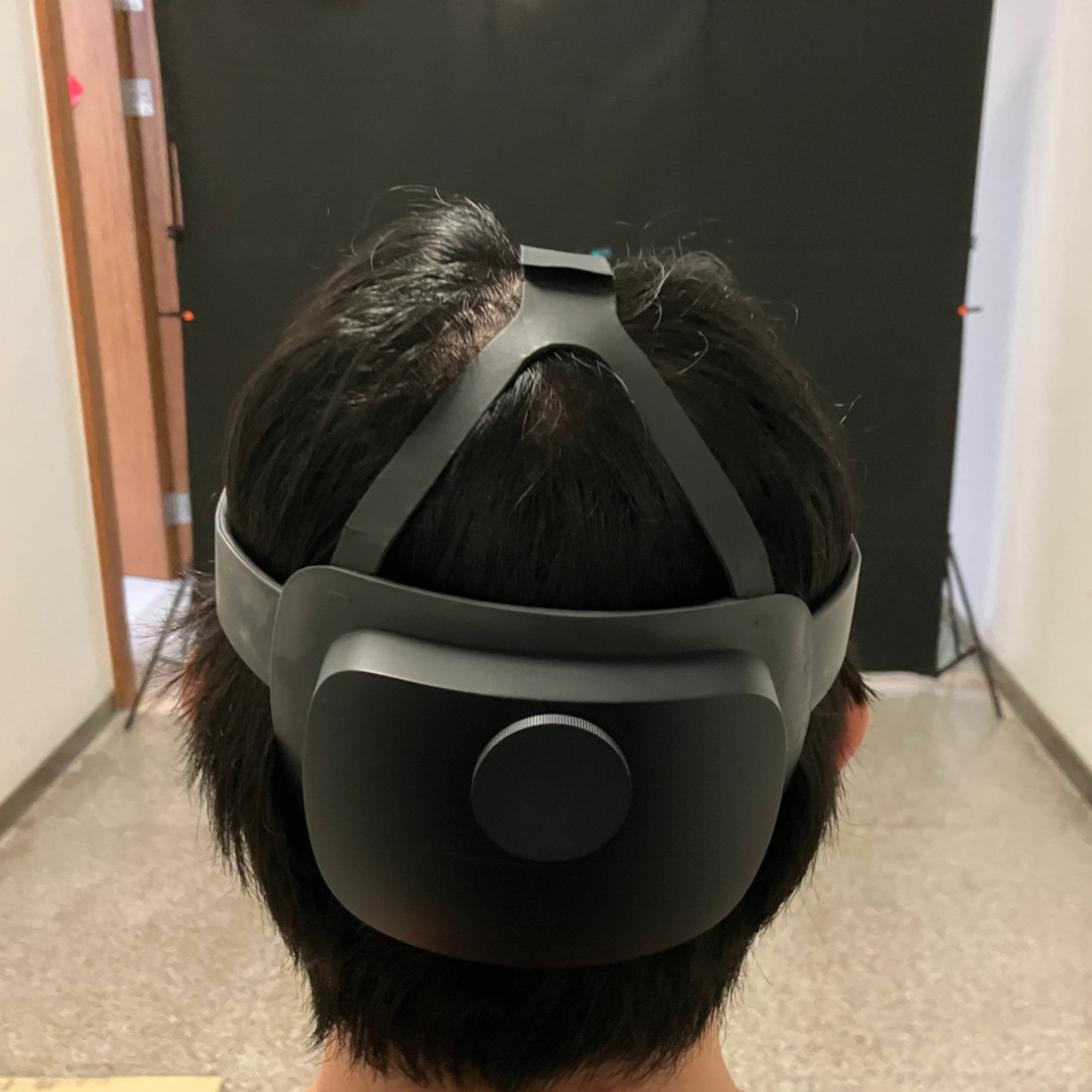}}
	\subfigure {\includegraphics[width=.22\textwidth]{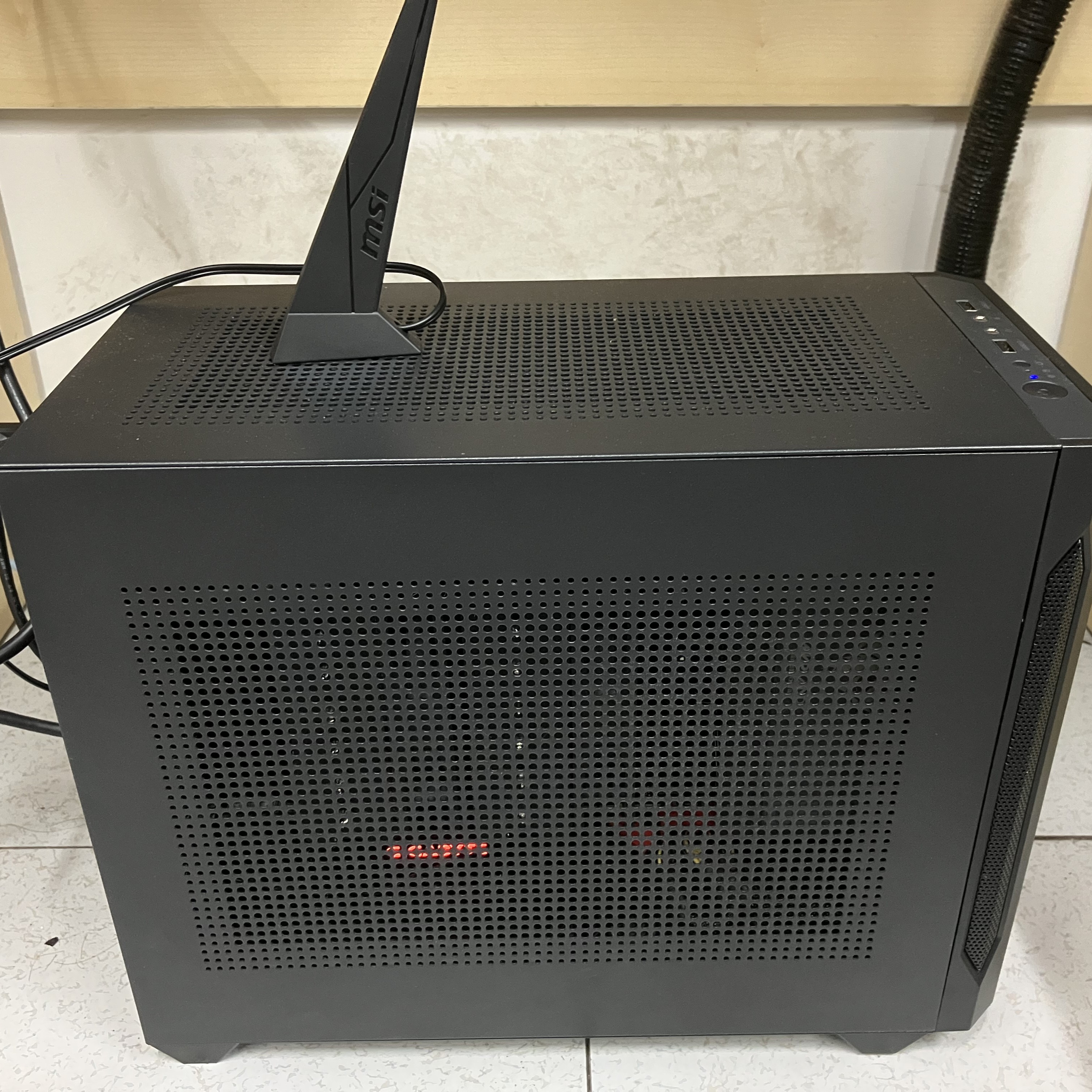}}
	\caption{Experiment settings.}
	\label{fig_3}
\end{figure}

We recruited 54 participants (29 males, 25 females; aged 19–25 years) for the gaze data collection. All participants reported normal or corrected-to-normal vision. Prior to the experiment, the eye tracker was calibrated individually for each participant. Participants performed the task in a dimly lit room while seated with minimal body movement and they were provided with a pair of earplugs to avoid auditory disturbance. During the experiment, participants were given a mandatory break after each round and could also request additional breaks if they felt tired or uncomfortable. A snapshot of the experimental setup is shown in \cref{fig_3}.

\subsection{Procedure}
Prior to the experiment, participants were given at least 10 minutes to familiarize themselves with the headset and the virtual environment to ensure task comprehension. Each participant then completed 6 rounds of fixation tasks, with each round consisting of 20 individual trials. In each trial, a start marker appeared at a random location within the 2 × 2-meter virtual area (fixed at 3 meters from the headset). Participants were required to maintain gaze fixation on this marker for 1 second to initiate the trial. Upon disappearance of the marker, a yellow arrow appeared at its former location, indicating the direction of the fixation target (a yellow cross). Participants were instructed to follow the arrow’s direction, locate the fixation target, and maintain steady gaze on its center for 5 seconds. After successful fixation, both arrow and target disappeared, triggering the next trial. Though conceptually divided into three phases—(1) trial initiation, (2) gaze shift, and (3) sustained fixation—the sequence flowed continuously without explicit pauses.

During each trial, we recorded the precise sampling time of each gaze point, the target position, the participant’s head position and direction, as well as the gaze origin position and gaze direction(measured in visual angles). Derived features, such as the linear velocity and angular velocity of gaze points, gaze position projected onto the target plane were subsequently computed from these measurements. To ensure data validity, trials were excluded if they met any of the following criteria: (1) Failure to maintain uninterrupted fixation on the target center for $\geq$ 2.5 seconds during the first 5-second time window of a trial; (2) Gaze shifts exceeding the boundaries of the 2×2-meter virtual area; (3) Loss of fixation, defined as $\geq$ 24 consecutive gaze points radially $>$ 0.1 m from the target center during the sustained fixation phase (corresponding to the 400 ms upper bound of the fixation window suggested in \cite{salvucci2000identifying}, and following HoloLens visual angle guidelines). In total, our collected data contains 54 participants’ exploration data in 3771 (54 × 20 × 6 - 2709) trials. Each trial data contains about 3,00 gaze points (60 Hz sampling rate) with features like target position, gaze position, gaze linear velocity, gaze angular velocity and timestamp information. Our dataset is named TimeGazer-dataset and is available online.

\section{Data Analysis}
\begin{figure}[h]
    \centering
    \includegraphics[width=0.35\textwidth]{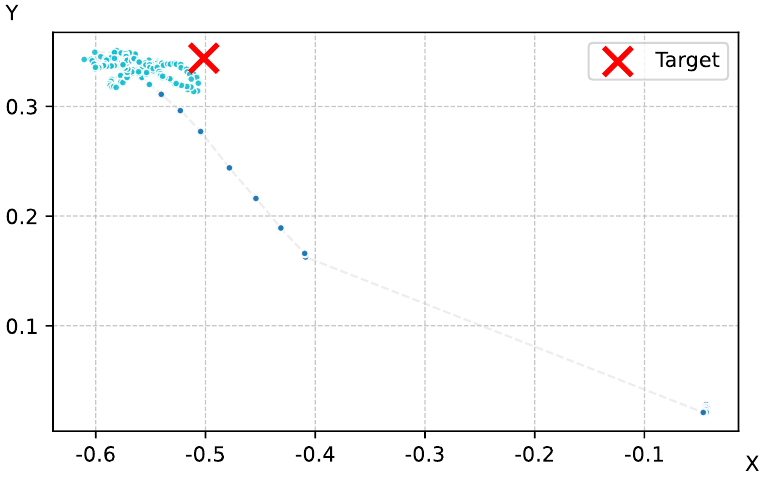}
    \caption{A typical gaze trajectory example.}
    \label{fig_4}
\end{figure}

The active gaze fixation task defined in this study is a composite visual task consisting of a target search phase guided by directional stimuli and a subsequent target-fixation phase. As illustrated in \cref{fig_4}, the eye-tracking data collected during this process can be categorized into two types: rapid movements and stable fixations (shown in different colors in the figure). Several prior studies have analyzed gaze movement characteristics in VR and leveraged them for future gaze prediction, such as \cite{hu2019sgaze,hu2020dgaze}. Unlikely, our research goal is to make gaze points during the stable fixation phase exhibit smaller overall deviations from the target and greater internal concentration.

\subsection{Analysis and Processing of Raw Gaze Data}\label{analysis_raw_data} In a given visual task, eye-movement data are commonly segmented into saccades and fixations using kinematic thresholds—typically speed or spatial dispersion—following principles established in physiology and classical eye-tracking research \cite{bach2012control,manor2003defining}. However, decades of studies since Yarbus et al. \cite{yarbus1967eye} have demonstrated that task instructions and stimulus characteristics profoundly alter eye-movement behavior, affecting fixation-duration distributions, saccade amplitudes, and scanpath patterns \cite{rothkopf2007task,borji2014defending}. Consequently, we adapted rather than directly adopted empirical thresholds, retaining the principles of classical I-DT and I-VT while tailoring the parameters to our active fixation setting\cite{andersson2017one,holmqvist2023retracted}. Specifically, the effective fixation region was defined as a circle of 0.1 m radius around the target, corresponding to approximately 3.81° of visual angle at our viewing distance(\cref{app:calculate_visual_angle}). Within this region, we analyzed the angular velocity distribution of gaze samples and observed that over 95\% of points fell below 15°/s (\cref{fig:app:angular_velocity}). This informed our threshold choice, which was further validated through sensitivity analyses (\cref{app:parameter_sensitivity_analysis}). We defined the onset of stable fixation as the first 12-sample window (200 ms at 60 Hz, following \cite{salvucci2000identifying}) within the target region where the mean angular velocity falls below 20°/s.

\subsection{Data Augmentation}\label{data_augmentation} However, despite careful thresholding, real fixation sequences often exhibit systematic offsets from the target and substantial within-fixation variability(\cref{fig_4}), which may introduce bias during model training. To mitigate this issue, we further generated synthetic fixation trajectories with reduced spatial bias and stronger internal cohesion. These synthetic sequences reflect the idealized convergence expected during the target-fixation phase, providing additional training data that mitigate the propagation of undesired offsets. The generation process and validation are detailed in \cref{app:synthetic_data_and_blending}.

% As shown in \cref{fig_4}, real fixation sequences during the target-fixation phase often exhibit systematic deviations from the target and considerable dispersion across samples. While such variability is natural in human behavior, it poses a significant challenge for our sequence-to-sequence prediction task, which requires consistent and target-centric future gaze trajectories. If left unaddressed, these biases would propagate through training, leading the model to reproduce undesired offsets rather than the expected convergence toward the target. 

% To address this issue, we designed a synthetic data generation and blending strategy that produces fixation sequences with reduced spatial deviations and stronger internal cohesion, reflecting the idealized attention distribution expected during the target-fixation phase. During training, real and synthetic gaze points are linearly interpolated to form mixed sequences, with the proportion of synthetic data gradually increased across folds. This curriculum-inspired strategy allows the model to first adapt to real distributions before progressively integrating synthetic variations, thereby enhancing stability and reducing overfitting risks \cite{bengio2009curriculum,wang2021survey}. The whole pipeline and validation of this strategy is elaborated on \cref{app:synthetic_data_and_blending}.

\section{TimeGazer Model}
\subsection{Problem Formulation}
Eye movement sequences during the target-fixation phase often exhibit irregular spatial dispersion and systematic offsets relative to the target center. To achieve gaze stabilization, we aim to transform these noisy fixation sequences into an idealized, target-centered trajectory.
Formally, let the historical gaze sequence captured during the target-searching phase be $\mathbf{X}_{1:T}=\{\mathbf{x}_1,\mathbf{x}_2,\cdots,\mathbf{x}_T\},\mathbf{x}_t\in \mathbb{R}^d$, where $\mathbf{x}_t$ represents the gaze feature vector (e.g., gaze position, velocity, and derived features) at time step $t$. The goal is to predict a future sequence $\mathbf{Y}_{T+1:T+\tau}=\{\mathbf{y}_{T+1},\mathbf{y}_{T+2},\cdots,\mathbf{y}_{T+\tau}\},\mathbf{y}_k\in \mathbb{R}^d$, where $\tau$ denotes the prediction horizon, and each $\mathbf{y}_k$ corresponds to an idealized fixation point that is both closer to the target and exhibits stronger spatial cohesion. The mapping can be expressed as a parameterized function: ${\tilde{\mathbf{Y}}_{T+1:T+\tau}} = f_\theta(\mathbf{X}_{1:T})$, where $f_\theta$ is learned by minimizing the discrepancy between predicted sequences and reference sequences (constructed via data augmentation, see \cref{data_augmentation}).

To capture temporal dependencies and sequence dynamics, we adopt a sequence-to-sequence (seq2seq) prediction framework, which naturally aligns with the task of predicting temporally coherent future gaze patterns from historical eye movements. We evaluate the model across multiple prediction horizons, reflecting varying levels of anticipatory gaze stabilization.

\subsection{Model Structure}\label{sec:model}
\begin{figure*}[!htbp]
    \centering
    \includegraphics[width=1.0\linewidth]{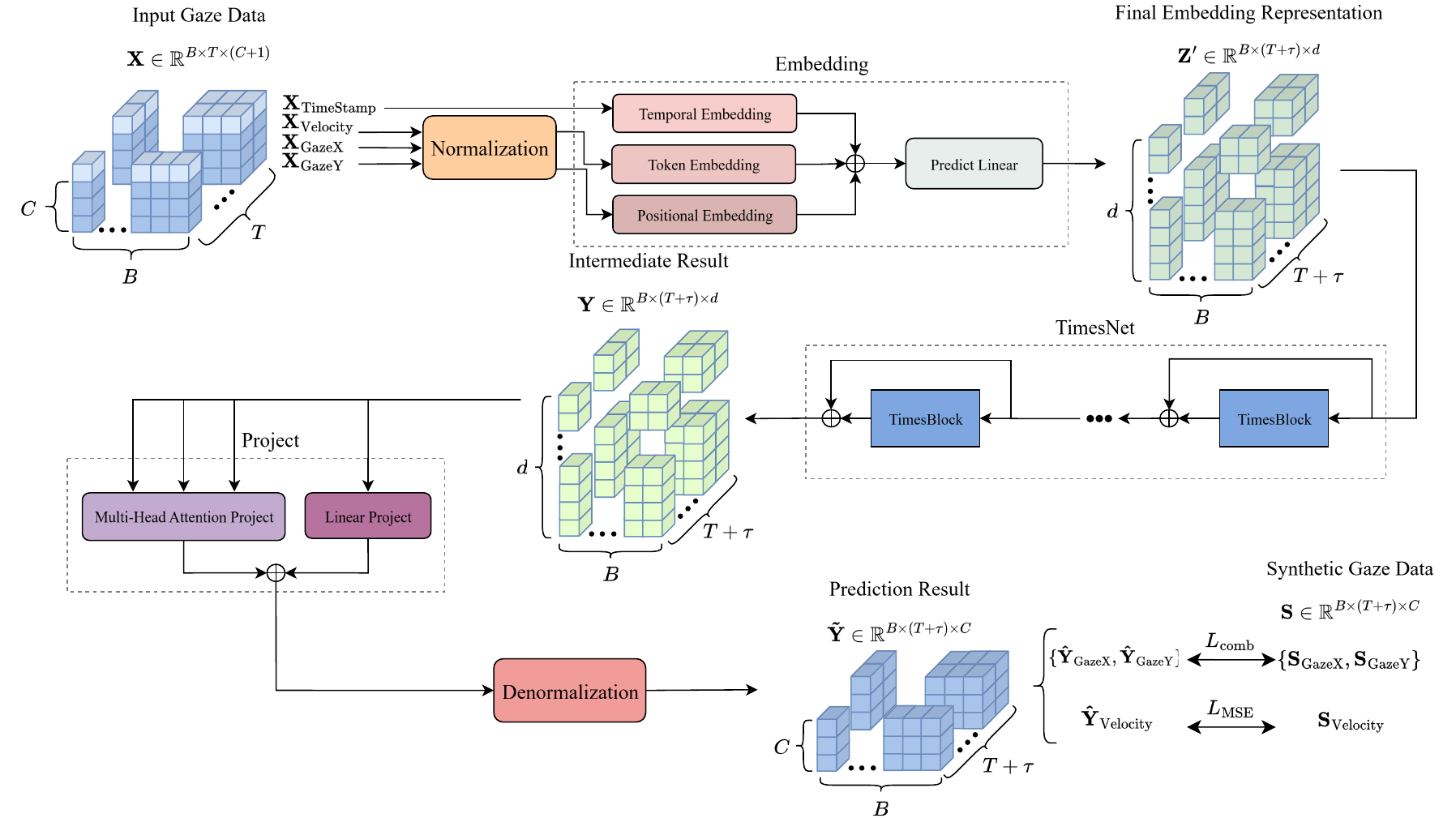}
    \vspace{-0.2in}
    \caption{Architecture of our TimeGazer model. Raw historical gaze sequences are first encoded via an embedding module, followed by temporal feature extraction and sequence-to-sequence modeling using TimesNet. The projected module then reduces the temporal representation to generate the final predicted gaze sequence.}
    \label{fig:timegazer_pipeline}
\end{figure*}

The overall architecture of TimeGazer is illustrated in \cref{fig:timegazer_pipeline}. The model input, $\mathbf{X}\in \mathbb{R}^{B\times T \times(C+1)}$, consists of gaze sequences represented as a tensor with batch size $B$, sequence length $T$, and $C$ gaze-related features, with an additional dimension corresponding to timestamp information.
\subsubsection{Embedding Module}
For each sample $b\in[1,B]$ and feature channel $c\in[1,C]$, gaze coordinates and velocities are standardized along the temporal dimension to ensure consistent input scales prior to embedding:
\begin{equation}
    \left\{
        \begin{array}{lr}
            \tilde{x}_{b,t,c}=\dfrac{x_{b,t,c}-\mu_{b,c}}{\sigma_{b,c}},\\
            \mu_{b,c}=\dfrac{1}{T}\sum_{t=1}^T x_{b,t,c}, \quad
            \sigma_{b,c}^2=\dfrac{1}{T}\sum_{t=1}^T (x_{b,t,c}-\mu_{b,c})^2.
        \end{array}
    \right.
\end{equation}
             
The normalized features $\tilde{\mathbf{X}}$ are then projected into an embedding space through a Token Embedding module, implemented as a 1D convolution layer with a kernel size of 3:
\begin{equation}
\mathbf{Z}_{\text{token}} = \text{Conv1D}_{k=3}(\tilde{\mathbf{X}}),
\quad \mathbf{Z}_{\text{token}} \in \mathbb{R}^{B \times T \times d}.
\end{equation}
This module not only projects features into a high-dimensional representation but also captures local temporal dependencies critical for sequence modeling. To incorporate global order information, we adopt sinusoidal positional encoding, which introduces positional context without additional trainable parameters. The encoding for each time step $t\in[1,T]$ is defined as:
\begin{equation}
    \left\{
        \begin{array}{lr}
            \mathbf{Z}{\text{position}}(t,2i) = \sin\left(\dfrac{t}{10000^{2i/d}}\right), \\
            \mathbf{Z}{\text{position}}(t,2i+1) = \cos\left(\dfrac{t}{10000^{2i/d}}\right),
        \end{array}
    \right.
\end{equation}
where d denotes the embedding dimension, and $i$ indexed the frequency components of the encoding. Specifically, $\mathbf{Z}\mathrm{position}(t,2i)$ and $\mathbf{Z}\mathrm{position}(t,2i+1)$ correspond to the $2i$-th and $(2i+1)$-th dimensions of the positional embedding at time step $t$. In parallel, timestamp features are embedded via a bias-free linear projection, yielding $\mathbf{Z}_{\text{time}}\in\mathbb{R}^{B\times T\times d}$. The final embedding representation is obtained through element-wise summation:
\begin{equation}
\mathbf{Z}=\mathbf{Z}_{\text{token}}+\mathbf{Z}_{\text{position}}+\mathbf{Z}_{\text{time}}, \quad\mathbf{Z}\in \mathbb{R}^{B\times T\times d}.
\end{equation}
Finally, $\mathbf{Z}$ is passed through a Predict Linear layer, which extends the temporal dimension from the historical sequence length $T$ to the combined length of $T+\tau$, where $\tau$ denotes the prediction horizon:
\begin{equation}
    \mathbf{Z}^{\prime}=\text{Linear}(\mathbf{Z}),\quad\mathbf{Z}^{\prime}\in \mathbb{R}^{B\times (T+\tau)\times d}.
\end{equation}

\subsubsection{Temporal Model}
To effectively capture the temporal dependencies in eye movements, we adopt TimesNet\cite{Wu_Hu_Liu_Zhou_Wang_Long_2023}, a state-of-the-art model for short-term temporal analysis, as the backbone of our framework. TimesNet is composed of a stack of TimesBlock modules connected via residual connections, enabling the model to jointly capture both short- and long-term dependencies in gaze sequential data. Within each TimesBlock, the input sequence is first transformed into the frequency domain using Fourier transform to extract periodic patterns of gaze points. Subsequently, multi-scale convolutional blocks are applied to capture local dependencies at different temporal resolutions. Finally, all periodic features are adaptively aggregated to form the intermediate temporal representation:
\begin{equation}
    \mathbf{Y}=\text{TimesNet}(\mathbf{Z}^{\prime}),\mathbf{Y}\in\mathbb{R}^{B\times {(T+\tau)} \times d},
\end{equation}
which serves as the basis for downstream prediction tasks.

\subsubsection{Project Layer}
To enhance the expressiveness of temporal representations while maintaining computational efficiency, we adopt two complementary projection strategies: (1) a multi-head self-attention projection, which captures long-range dependencies by dynamically weighting temporal positions, and (2) a linear projection, which provides a lightweight mapping that preserves local structures.

The attention-based projection is defined as:
\begin{equation}
\mathbf{\hat{Y}}_{\text{attn}} = \text{MHA}(\mathbf{Y}),\quad \mathbf{\hat{Y}}_{\text{attn}}\in\mathbb{R}^{B\times (T+\tau)\times C},
\end{equation}
where $\text{MHA}(\cdot)$ is the Multi-Head Attention. 

The linear projection is defined as:
$\mathbf{\hat{Y}}_{\text{linear}}\in \mathbb{R}^{B\times {(T+\tau)\times C}}$ is defined as:
\begin{equation}
\mathbf{\hat{Y}}_{\text{linear}} = \mathbf{Y}\mathbf{W}_{\text{linear}} + \mathbf{b}_{\text{linear}}, \quad
\mathbf{W}_{\text{linear}}\in \mathbb{R}^{d\times C}, \mathbf{b}_{\text{linear}}\in \mathbb{R}^{C}.
\end{equation}
Where $\mathbf{W}_{\text{linear}}$ and $\mathbf{b}_{\text{linear}}$ are trainable parameters.

Finally, the fused projection combines the two branches through a learnable balance parameter $\alpha$:
\begin{equation}
\mathbf{\hat{Y}} = \alpha \mathbf{\hat{Y}}_{\text{attn}} + (1-\alpha)\mathbf{\hat{Y}}_{\text{linear}}, \quad \alpha \in [0,1].
\end{equation}

We also performed ablation experiments to compare the two single-branch variants with the fused design (\cref{sec:ablation_study}). In addition, We conducted a sensitivity analysis on the number of attention heads used in the Multi-Head Attention (MHA) Project module, with the embedding dimension fixed at $d=16$. As shown in \cref{tab:hyper_attention}, the model achieves the best performance when the number of heads is set to 8, while performance drops noticeably when the number of heads increases to 16. This decline can be attributed to the fact that with $d=16$ and $h=16$, each attention head only has one dimension, which severely limits its representational capacity and hinders effective information interaction across dimensions. These results suggest that an excessively large number of heads may be detrimental under low embedding dimensions, highlighting the importance of balancing head number and embedding size in practice.
\begin{table}[!htbp]
    \centering
    \caption{Hyperparameter sensitivity analysis on the number of attention heads in the MHA Project (embedding dimension $d=16$). Values are reported as mean $\pm$ standard deviation (SD).}
\begin{tabular}{cccc}
\hline
Head Number & CI $\uparrow$    & AI $\uparrow$   & AD $\downarrow$   \\ \hline
2           & $6.30 \pm 10.78$ & $1.38 \pm 1.01$ & $0.051 \pm 0.059$ \\
4           & $6.33 \pm 12.12$ & $1.41 \pm 1.04$ & $0.050 \pm 0.060$ \\
8           & $\textbf{6.40} \pm 11.73$ & $\textbf{1.41} \pm 0.99$ & $\textbf{0.050} \pm 0.059$ \\
16          & $4.46 \pm 8.34$  & $1.40 \pm 0.94$ & $0.050 \pm 0.061$ \\ \hline
\end{tabular}
    \label{tab:hyper_attention}
\end{table} 
%(\cref{app:attention_head_number}). 
\subsection{Loss Function}
We design a loss function that not only penalizes pointwise prediction errors but also explicitly constrains the spatial distribution of predicted gaze points. Specifically, our formulation extends the conventional Mean Squared Error (MSE) by introducing center distance and dispersion consistency regularization terms.

Formally, given a set of predicted fixation points $\hat{\mathbf{P}}\{\hat{p}_1,\cdots,\hat{p}_n\}$ and ground-truth fixation points ${\mathbf{P}}\{{p}_1,\cdots,{p}_n\}$, the per-step loss for the fixation phase is defined as:
\begin{equation}
    \begin{aligned}
        &{L}_\text{comb} = \\
        &\quad\underbrace{\frac{1}{n}\sum_{i=1}^n ||\hat{p}_i - p_i||_2^2}_{\text{MSE loss}}+
        \lambda_c \underbrace{||\mu(\hat{\mathbf{P}})-\mu(\mathbf{P})||_2^2}_{\text{center distance}}+
        \lambda_v \underbrace{||\sigma^2(\hat{\mathbf{P}})-\sigma^2(\mathbf{P})||_1}_{\text{dispersion consistency}},
    \end{aligned}
\end{equation}
where $\mu(\cdot)$ and $\sigma^2(\cdot)$ denote the empirical mean and variance of the point set, and $\lambda_c$ and $\lambda_v$ are the balancing weights. The first term (MSE) ensures pointwise closeness to the ground truth, the center distance prevents global bias in predictions, and the dispersion consistency preserves the variance of the predicted points, avoiding over-collapse.

To further investigate the role of each component in the combined loss, we first estimated the relative scales of the three terms by computing their average magnitudes on small batches, and accordingly set $\lambda_c$ and $\lambda_v$ to be within comparable orders. Then, we performed a coarse grid search and found the best-performing configuration to be $\lambda_c=0.001$ and $\lambda_v=0.05$. 
\begin{table}[!htbp]
    \centering
    \caption{Ablation analysis of the loss weight $\lambda_c$ and $\lambda_v$. Values are reported as mean $\pm$ standard deviation (SD).}
\begin{tabular}{clccc}
\hline
$\lambda_c$ & \multicolumn{1}{c}{$\lambda_v$} & CI $\uparrow$    & AI $\uparrow$   & AD $\downarrow$   \\ \hline
\textemdash          & \textemdash            & $ 6.23\pm11.53 $ & $ 1.42\pm1.17 $ & $ 0.050\pm0.060 $ \\
\textemdash          & \checkmark             & $ 6.41\pm11.53 $ & $1.38 \pm0.99 $ & $ 0.051\pm0.059 $ \\
\checkmark           & \textemdash            & $ 6.22\pm11.53 $ & $ 1.43\pm1.08 $ & $ 0.050\pm0.060 $ \\
\checkmark           & \checkmark             & $\textbf{6.40}\pm11.73 $  & $ \textbf{1.43}\pm0.99 $ & $ \textbf{0.050}\pm0.059 $ \\ \hline
\end{tabular}
    \label{tab:loss_component}
\end{table}

Finally, we ablated the individual components of $L_{\text{comb}}$, as shown in \cref{tab:loss_component}, all the metrics used in the validation are defined in \cref{sec:metrics}. Removing either the center distance or the dispersion consistency term resulted in performance drops across AI, CI, and AD, demonstrating that both terms make complementary contributions. The best results were consistently achieved when both components were included, confirming the necessity of the complete $L_{\text{comb}}$ design.

The overall loss for training is formulated as:
\begin{equation}
    L=\lambda L_\text{comb}+(1-\lambda)L_\text{velocity}+L_2^\text{reg}
\end{equation}
where $L_\text{velocity}$ computes the MSE of velocity between consecutive points over the entire sequence, and $L_2^\text{reg}$ is a standard weight decay to prevent overfitting.

To validate the necessity of each component, we conducted ablation studies by progressively enabling the center and variance constraints in addition to the baseline MSE(\cref{sec:ablation_study}). We also performed a hyperparameter sensitivity analysis on the weighting parameter $\lambda$ to assess its impact on model performance. We analyzed the sensitivity of the weighting factor $\lambda$ in the total loss:
\[
L = \lambda L_{\text{comb}} + (1-\lambda) L_{\text{velocity}} + L_2^{\text{reg}} .
\]
As shown in \cref{tab:hyper_loss}, increasing the weight of $L_{\text{comb}}$ consistently improved performance across all metrics. Larger $\lambda$ values led to higher AI scores, along with gradual improvements in CI and AD. The best overall performance was observed when $\lambda$ was set to 0.9. Therefore, we adopted $\lambda=0.9$ as the default setting in all subsequent experiments.

\begin{table}[!h]
    \centering
    \caption{Sensitivity analysis of the loss weight $\lambda$. Higher AI, lower CI and AD indicate better performance. Values are reported as mean $\pm$ standard deviation (SD).}
\begin{tabular}{llll}
\hline
$\lambda$   & AI$\uparrow$            & CI$\uparrow$            & AD$\downarrow$   \\ \hline
0.1 & 2.86±5.38  & 1.29±0.88  & 0.054±0.063 \\
0.3 & 4.61±8.23  & 1.35±0.95  & 0.051±0.061 \\
0.5 & 5.21±9.79  & 1.33±0.92  & 0.050±0.060 \\
0.7 & \underline{6.02}±10.79 & \underline{1.40}±0.98 & \underline{0.050}±0.060 \\
0.9 & \textbf{6.41}±11.73    & \textbf{1.41}±0.99    & \textbf{0.050}±0.059 \\ \hline
\end{tabular}
    \label{tab:hyper_loss}
\end{table}
% (\cref{app:hyper_loss}).

\subsection{Training Strategy}
\begin{figure}[!htbp]
    \centering
    \includegraphics[width=0.8\linewidth]{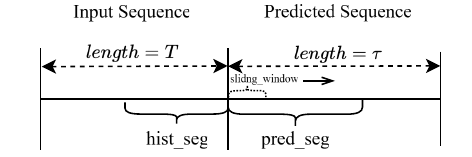}
    \caption{Sliding window training strategy.}
    \label{fig:sliding_window}
\end{figure}

Training seq2seq models with the conventional teacher forcing strategy often leads to exposure bias, since during inference the model must rely on its own predictions rather than ground-truth inputs, causing error accumulation over time \cite{bengio2015scheduled}. To address this issue, we draw inspiration from SCINet \cite{liu2022scinet} and adopt a sliding-window training strategy. As shown in \cref{fig:sliding_window}, both the historical and predicted sequence lengths are fixed. At each step, the model generates a predicted segment that is longer than the sliding-window size. We then take only the first $l_\text{w}$ points of this segment, where $l_\text{w}$ is the sliding-window length, append them to the growing predicted sequence, and shift the historical window forward by $l_\text{w}$ points to form the next input. This process repeats until the predicted sequence reaches the horizon $\tau$. By partially consuming the model’s own predictions in a progressive manner, the strategy reduces reliance on ground-truth samples, thereby alleviating exposure bias and improving long-term robustness. 

We further investigated the effect of the sliding-window size on model performance, all the metrics used in the validation are defined in \cref{sec:metrics}. As shown in \cref{tab:sliding_window}, the model achieves its best results when the window size is set to 16. A larger window provides the model with a broader temporal context, which facilitates more accurate gaze prediction. Based on this observation, we adopt a window size of 16 as the default configuration in all subsequent experiments.
\begin{table}[!hb]
\centering
\caption{Hyperparameter sensitivity analysis on sliding window size. Values are reported as mean $\pm$ standard deviation (SD).}
\begin{tabular}{ccccc}
\hline
Window Size & CI $\uparrow$ & AI $\uparrow$ & AD $\downarrow$ \\ \hline
4           & $ 6.03 \pm 9.94 $ & $ 1.38 \pm 0.95 $ & $ 0.051 \pm 0.059 $  \\
8           & $ 5.36 \pm 9.06 $ & $ 1.38 \pm 1.02 $ & $ 0.050 \pm 0.060 $  \\
16          & $\textbf{6.40} \pm 11.73$ & $\textbf{1.41} \pm 0.99$ & $\textbf{0.050} \pm 0.059$ \\ \hline
\end{tabular}
    \label{tab:sliding_window}
\end{table}

\section{Experiments and Results}
To comprehensively evaluate the effectiveness of our model, we conducted a series of experiments. First, we examined its adaptability to different prediction horizons. Second, we implemented a gaze-selection task in Unity and performed cross-user experiments to assess real-world performance on AR devices. Third, we carried out ablation studies to validate the contributions of individual model components. Finally, we conducted parameter sensitivity analyses to evaluate the robustness of the model with respect to key hyperparameters.
\subsection{Metrics}\label{sec:metrics}
We adopt two relative metrics and one absolute metric to evaluate gaze stabilization for each gaze sequence with $n_i$ points.

1. Concentration Improvement (CI):
Concentration is measured by the standard deviation of distances between gaze points and targets:
\begin{equation}
    \text{CI}=\frac{\sqrt{\frac{1}{n_i}\sum^{n_i}_{j=1}||\mathbf{p}_{j}-\mathbf{g_i}||^2_2}}{\sqrt{\frac{1}{n_i}\sum^{n_i}_{j=1}||\mathbf{\hat{p}}_{j}-\mathbf{g_i}||^2_2}+\epsilon},
\end{equation}
where $\mathbf{g}_i\in \mathbb{R}^2$ denotes the coordinate of target point, $\mathbf{p}_{j}\in \mathbb{R}^2$ and $\mathbf{\hat{p}}_{j}^2$ are predicted point and origin point, respectively, and $\epsilon$ is a minimum value.

2. Accuracy Improvement (AI):
Accuracy is measured by the mean distance $\bar{d_i}$ between $n_i$ points and the target. Similar to CI, AI is defined as:
\begin{equation}
    \text{AI}=\frac{\frac{1}{n_i}\sum^{n_i}_{j=1}||\mathbf{p}_{j}-\mathbf{g_i}||_2}{\frac{1}{n_i}\sum^{n_i}_{j=1}||\mathbf{\hat{p}}_{j}-\mathbf{g_i}||_2+\epsilon}.
\end{equation}

3.Average Distance (AD):
AD is the absolute mean distance of the predicted points to the target:
\begin{equation}
    \text{AD}=\frac{1}{n_i}\sum^{n_i}_{j=1}||\mathbf{\hat{p}}_{j}-\mathbf{g_i}||_2.
\end{equation}

\subsection{Experimetn Settings}
Our model is implemented in PyTorch and optimized using Adam with an initial learning rate of 0.001. A cosine annealing scheduler is applied: $\eta_{t}=\frac{\eta_0}{2}(1+\cos(\frac{\pi \cdot \text{epoch}_t}{E})), t\in[1,E]$, where $\eta_0$ is the initial learning rate and $E$ the total number of training epochs. Early stopping with patience of 20 is used to prevent overfitting. We set the length of the historical gaze sequence to 96 time steps (\cref{app:define_historical_sequence_length}) and tested different batch sizes in theoretical experiments and adopted 64 for user studies, balancing accuracy and efficiency. For the sliding-window strategy, the input sequence length is fixed at $T=64$ with a stride of 16. Various prediction horizons $\tau$ were explored in theoretical experiments, while $\tau=64$ was selected for user studies. 

\subsection{Quantitative Evaluation}
Before the user study, we performed quantitative experiments to evaluate TimeGazer under varying prediction horizons and batch sizes.
\begin{table}[!h]
\centering
\caption{Theoretical evaluation results. Values are reported as mean $\pm$ standard deviation (SD).}
\setlength{\tabcolsep}{3pt} % 缩小列间距
\small % 缩小整体字体
\begin{tabular}{c c c c c c}
\hline
\begin{tabular}[c]{@{}c@{}}Predicted\\ length\end{tabular} & Batchsize & AI$\uparrow$ & CI$\uparrow$ & AD$\downarrow$ & \begin{tabular}[c]{@{}c@{}}Time cost\\ per round(s)$\downarrow$\end{tabular} \\ \hline
\multirow{3}{*}{32($\approx500ms$)}                                        & 16        & 3.17$\pm$9.84   & 1.33$\pm$0.94 & 0.049$\pm$0.054 & 0.030$\pm$0.0047                                                     \\
                                                           & 32        & 3.44$\pm$5.35   & 1.44$\pm$1.05 & 0.048$\pm$0.055 & 0.029$\pm$0.0015                                                     \\
                                                           & 64        & 2.88$\pm$6.76   & 1.36$\pm$1.01 & 0.050$\pm$0.057 & 0.031$\pm$0.0036                                                     \\ \hline
\multirow{3}{*}{64($\approx1000ms$)}                                        & 16        & 6.16$\pm$12.2   & 1.36$\pm$0.98 & 0.051$\pm$0.058 & 0.058$\pm$0.0057                                                     \\
                                                           & 32        & 6.60$\pm$23.31  & 1.37$\pm$0.98 & 0.051$\pm$0.061 & 0.058$\pm$0.0057                                                     \\
                                                           & 64        & 6.41$\pm$11.73  & 1.42$\pm$0.99 & 0.049$\pm$0.059 & 0.055$\pm$0.0046                                                     \\ \hline
\multirow{3}{*}{96($\approx1500ms$)}                                        & 16        & 9.77$\pm$16.33  & 1.41$\pm$1.10 & 0.051$\pm$0.059 & 0.082$\pm$0.0094                                                     \\
                                                           & 32        & 7.25$\pm$12.12  & 1.36$\pm$0.96 & 0.051$\pm$0.060 & 0.079$\pm$0.0011                                                     \\
                                                           & 64        & 6.20$\pm$10.47  & 1.42$\pm$1.15 & 0.050$\pm$0.061 & 0.082$\pm$0.0036                                                     \\ \hline
\multirow{3}{*}{128($\approx 2000ms$)}                                       & 16        & 9.87$\pm$15.84  & 1.42$\pm$1.24 & 0.051$\pm$0.060 & 0.11$\pm$0.0090                                                      \\
                                                           & 32        & 12.84$\pm$22.66 & 1.44$\pm$1.27 & 0.051$\pm$0.059 & 0.11$\pm$0.0051                                                      \\
                                                           & 64        & 6.89$\pm$11.20  & 1.37$\pm$1.06 & 0.057$\pm$0.072 & 0.11$\pm$0.0066                                                      \\ \hline
\end{tabular}
    \label{tab:theoratical_exp}
\end{table}
As shown in \cref{tab:theoratical_exp}, increasing the prediction length improves the AI metric, indicating more concentrated gaze predictions, but also leads to a substantial rise in computational cost per prediction. Moreover, the consistently low Average Deviation (AD) values across different participants—on the order of $10^{-2}$ to $10^{-3}$—demonstrate that the learned temporal patterns generalize well, making the model robust and applicable across individuals despite inter-subject variability in gaze behavior. Balancing performance and efficiency, we adopt a prediction length of 64 (approximately one second) and a batch size of 64 for the subsequent user evaluation.

\subsection{User Study and Performance Evaluation}
To evaluate the performance of TimeGazer in AR scenarios, we conducted a user study with 27 participants (16 male, 11 female; aged 18–25 years, $M=21$). Prior to the experiment, each participant underwent a calibration procedure to ensure accurate eye-tracking on the device and completed a short training session to familiarize themselves with the task. Our method adopted counterbalanced and randomized ordering of the two algorithm conditions being compared, to mitigate potential sequence and learning effects.

\begin{figure}[!htb]
    \centering
    % 上方用户实验场景
    \subfigure[User study settings.]{
        \includegraphics[width=0.28\linewidth]{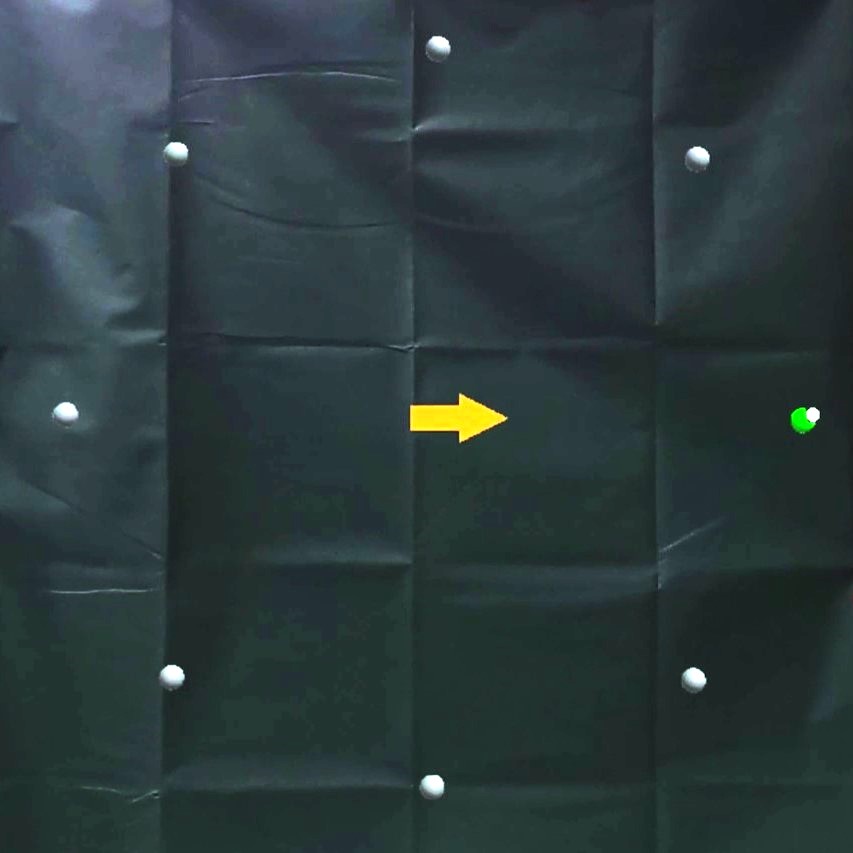}
        \label{fig:user_study}
    }
    % 下方两个MRTK子场景
    \subfigure[Target selection scenario.]{
        \includegraphics[width=0.28\linewidth]{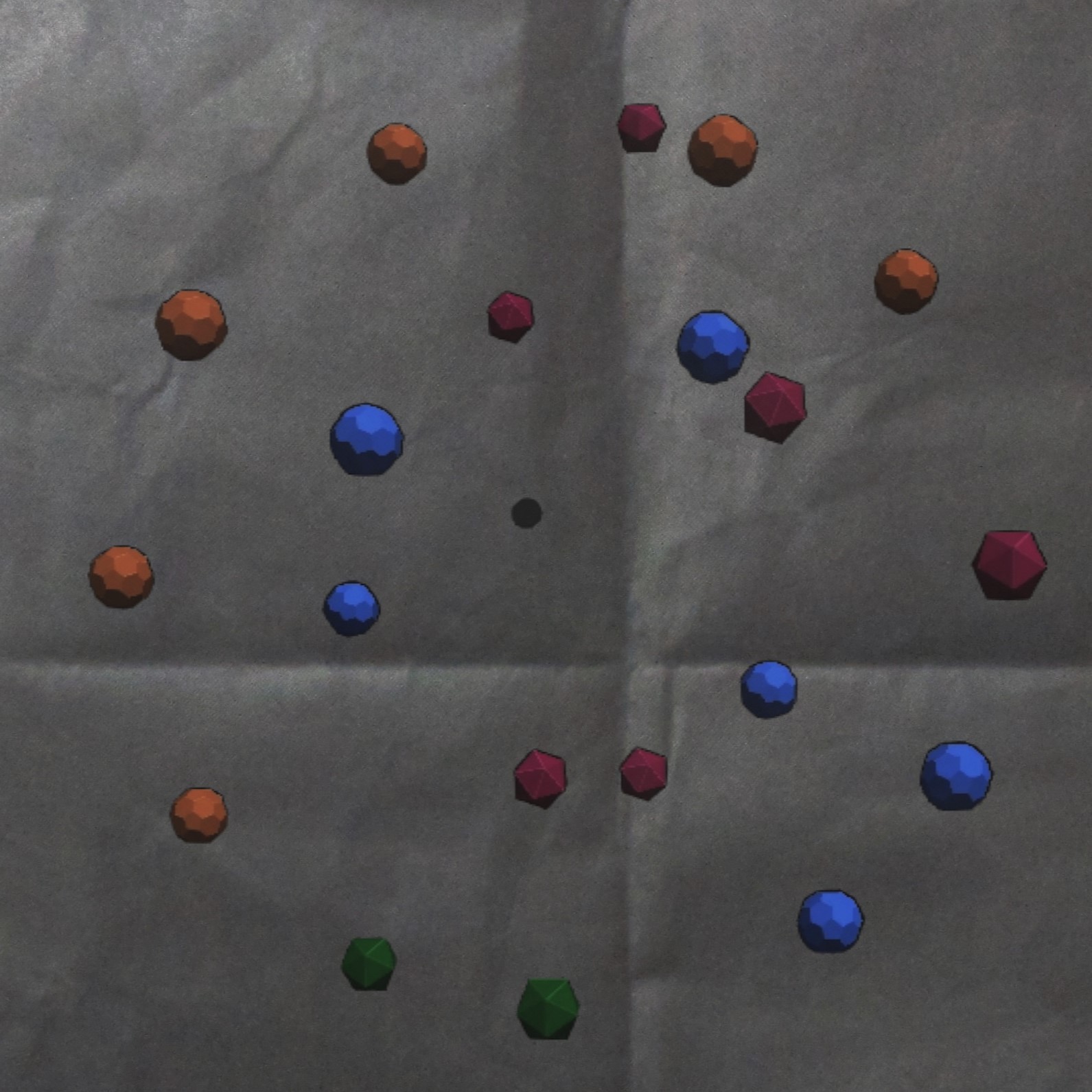}
        \label{fig:mrtk_selection}
    }
    % \hspace{6pt}
    \subfigure[Navigation scenario.]{
        \includegraphics[width=0.28\linewidth]{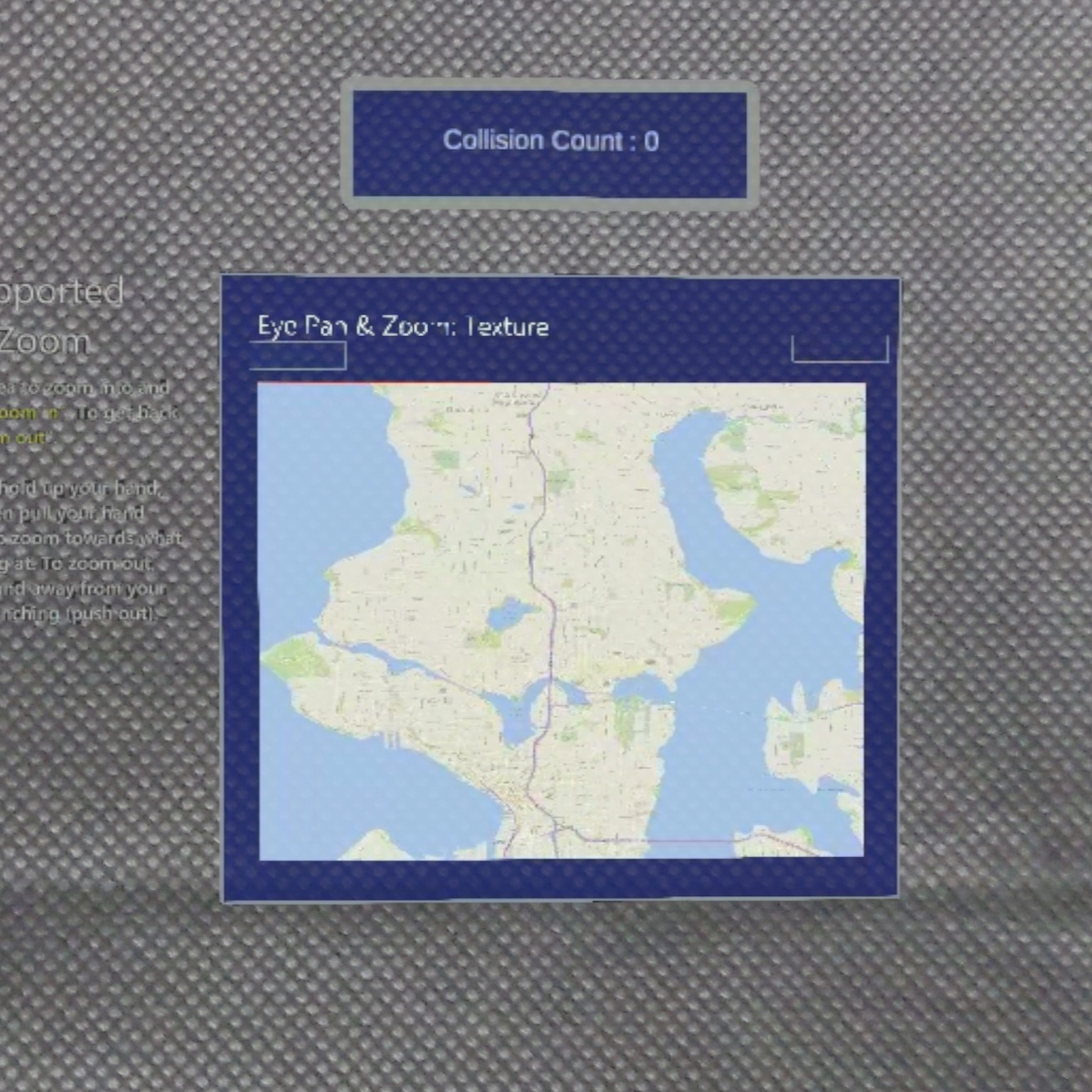}
        \label{fig:mrtk_navigation}
    }
    \caption{User study and MRTK-based evaluation scenarios: (a) controlled AR fixation task, (b) target selection, and (c) navigation.}
    \label{fig:user_study_all}
\end{figure}

We developed a gaze selection task in Unity, consisting of spherical targets with three radii: 0.10 m, 0.06 m, and 0.04 m. Each condition was repeated three times. As illustrated in \cref{fig:user_study}, during each trial participants were guided by directional arrows to sequentially fixate on eight target spheres. A fixation was considered successful if the participant maintained continuous gaze on a sphere for at least 1 s within a 5 s time limit, upon which the sphere was removed.

For each target, we recorded the target coordinates, participants’ gaze trajectories, fixation completion status, the number of fixation interruptions, the cumulative fixation duration, and the total task completion time.

We compared TimeGazer with the native HoloLens 2 eye-tracking algorithm using four metrics: (1) average fixation completion rate (ACR, ↑, proportion), (2) average task duration (ATD, ↓, seconds), (3) gaze interruptions per successful trial (GI, ↓, count/trial), and (4) fixation-to-threshold ratio per failed trial (FTR, ↑, proportion). Group means $\pm$ standard deviations, paired $p$-values, and Cohen’s $d$ are reported in \cref{tab:user_study}, \cref{tab:overall_results} and \cref{tab:user_survey}. Statistical significance is indicated as: *$p<0.05$, **$p<0.01$, ***$p<0.001$. Symbols ↑ and ↓ denote whether higher or lower values indicate better performance, respectively.
\begin{table}[!h]
\setlength{\tabcolsep}{2pt}
\centering
\footnotesize
\caption{Comparison of TimeGazer and HoloLens 2 native gaze tracking across four performance metrics. Values are reported as mean $\pm$ SD.}
\begin{tabular}{lccll}
\hline
Performance & TimeGazer & HoloLens 2 Native & $p$-value & Effect size \\ \hline
ACR(–) $\uparrow$        & $0.60 \pm 0.23$ & $0.51 \pm 0.25$ & 0.0034$^{**}$ & 0.873 \\
ATD(s) $\downarrow$      & $30.58 \pm 5.04$ & $32.56 \pm 5.14$ & 0.035$^{*}$ & 0.917 \\
GI(count/trial) $\downarrow$ & $1.87 \pm 0.48$ & $2.14 \pm 0.71$ & 0.035$^{*}$ & 0.397 \\
FTR(–) $\uparrow$        & $0.26 \pm 0.10$ & $0.23 \pm 0.084$ & 0.035$^{*}$ & 0.416 \\ \hline
\end{tabular}
\label{tab:user_study}
\end{table} 

As shown in \cref{tab:user_study}, TimeGazer outperformed the native HoloLens 2 on all four metrics, with statistically significant improvements (ACR: $p=0.0034$, $d=0.873$; ATD: $p=0.035$, $d=0.917$; GI: $p=0.035$, $d=0.397$; FTR: $p=0.035$, $d=0.416$), demonstrating its substantial impact on gaze stability and interaction efficiency.

In addition, we further evaluated TimeGazer in two standard AR interaction scenarios provided by the MRTK Eye Tracking library: \textit{Target Selection} and \textit{Navigation} (as illustrated in \cref{fig:mrtk_selection} and \cref{fig:mrtk_navigation}). For the \textit{Target Selection} task, we measured ATD to quantify the average time required to select a target, and GI to capture how often the gaze left the selection area during each trial. For the \textit{Navigation} task, we used ATD to evaluate the time required for users to follow a predefined path using gaze input. 

\begin{table}[!htbp]
\footnotesize
\centering
\caption{Performance comparison between TimeGazer and HoloLens 2 Native in MRTK scenarios. Values are reported as mean $\pm$ SD.}
\label{tab:overall_results}
\subtable[Selection]{
    \setlength{\tabcolsep}{3pt} 
    \renewcommand{\arraystretch}{0.9} 
    \begin{tabular}{@{}lccll@{}} 
    \hline
    Performance                  & TimeGazer & HoloLens 2 Native & $p$-value & Effect size \\
    ATD(s) $\downarrow$          & $ 1.77\pm 0.24$& $ 2.29\pm 0.56$& 0.0011$^{**}$& 1.26\\
    GI(count/trial) $\downarrow$ & $1.11 \pm 0.06 $& $ 1.17\pm0.10 $& 0.0482$^{*}$& 0.64\\
    \hline
    \end{tabular}
} 
\subtable[Navigation]{
    \setlength{\tabcolsep}{3pt}
    \renewcommand{\arraystretch}{0.9}
    \begin{tabular}{@{}lccll@{}}
    \hline
    Performance         & TimeGazer & HoloLens 2 Native & $p$-value & Effect size \\
    ATD(s) $\downarrow$ & $ 15.74\pm 3.08$& $ 19.05\pm 4.34$& 0.0008$^{***}$& 1.99\\
    \hline
    \end{tabular}
}
\end{table}

As shown in \cref{tab:overall_results}, TimeGazer outperformed the native HoloLens 2 eye-tracking algorithm across all MRTK scenarios, with statistically significant improvements in both target selection (ATD: $p=0.0011$, $d=1.26$; GI: $p=0.0482$, $d=0.64$) and navigation (ATD: $p=0.0008$, $d=1.99$). These results demonstrate that TimeGazer effectively enhances gaze-based interaction efficiency and stability in AR environments.

\begin{table}[!htbp]
\setlength{\tabcolsep}{3pt}
\footnotesize
    \centering
    \caption{Subjective ratings (5-point Likert scale) of stability and sensitivity for TimeGazer and the HoloLens 2 native eye-tracking algorithm. Higher values indicate better performance. $p$-values are from Wilcoxon signed-rank tests.}
\begin{tabular}{lcccc}
\hline
Metric      & TimeGazer  & HoloLens 2 Native & $p$-value &Effect size\\ \hline
Stability$\uparrow$   & 3.93$\pm$ 0.42           & 3.27$\pm$ 0.46           &0.0056$^{**}$   &1.234        \\
Sensitivity$\uparrow$ & 4.07 $\pm$ 0.46          & 3.07 $\pm$ 0.46           &0.0003$^{***}$  &4.009      \\ \hline
\end{tabular}
    \label{tab:user_survey}
\end{table}
To further evaluate user experience, participants provided 5-point Likert scale ratings for both methods after completing all trials for each target size. Ratings assessed two aspects of gaze selection: stability and sensitivity. Statistical significance was tested using the Wilcoxon signed-rank test, with results shown in \cref{tab:user_survey}. Users rated TimeGazer significantly higher than the native HoloLens 2 algorithm in both stability (3.93 $\pm$ 0.42 vs. 3.27 $\pm$ 0.46, $p=0.0056$, $d=1.234$) and sensitivity (4.07 $\pm$ 0.46 vs. 3.07 $\pm$ 0.46, $p=0.0003$, $d=4.009$), indicating that TimeGazer provides a noticeably more stable and responsive gaze interaction experience.

\subsection{Ablation Study}\label{sec:ablation_study}
We conducted ablation studies on the architectural design of TimeGazer to verify the effectiveness of each module.

First, we examined the components of the Embedding Module. As shown in \cref{tab:ablation_embed}, removing any of the three embedding types (token, position, or time) substantially degrades overall performance, indicating that each embedding plays an indispensable role in capturing different aspects of gaze dynamics.
\begin{table}[!htbp]
\setlength{\tabcolsep}{3pt}
    \centering
    % \footnotesize{46}
    \caption{Ablation study on embedding module. Values are reported as mean $\pm$ standard deviation (SD).}
\begin{tabular}{cccccc}
\hline
Token& Position& Time& CI $\uparrow$& AI $\uparrow$& AD $\downarrow$      \\ \hline
\textemdash & \checkmark  & \checkmark      & $4.62 \pm 8.60$      & $1.30 \pm 0.88$      & $0.053 \pm 0.061$    \\
\checkmark  & \textemdash & \checkmark      & $4.79 \pm 10.67$    & $1.35 \pm 0.88$      & $0.050 \pm 0.059$    \\
\checkmark  & \checkmark  & \textemdash     & $4.96 \pm 9.69$     & $1.38 \pm 0.97$      & $0.050 \pm 0.061$    \\ 
\checkmark  & \checkmark  & \checkmark      & $\textbf{6.41} \pm 11.73$   & $\textbf{1.41} \pm 0.99$    
 & $\textbf{0.050} \pm 0.059$    \\ \hline
\end{tabular}
    \label{tab:ablation_embed}
\end{table}

Next, we ablated the Project Module, with results summarized in \cref{tab:ablation_project}. Using only the Linear Project yields relatively high standard deviations, suggesting weaker robustness against outliers. By contrast, using only the Multi-Head Attention Project provides more stable predictions but offers limited improvement in point clustering compared to the Linear Project. Importantly, combining both components achieves the best balance: it enhances clustering (CI) and accuracy (AI) simultaneously while maintaining low variance
\begin{table}[!h]
    \centering
    \caption{Ablation study on project module. Values are reported as mean $\pm$ standard deviation (SD).}
\begin{tabular}{ccccc}
\hline
MHA & Linear & CI $\uparrow$ & AI $\uparrow$ & AD $\downarrow$ \\ \hline
\checkmark & \textemdash & $5.17 \pm 9.26$ & $1.35 \pm 1.08$ & $0.051 \pm 0.057$ \\
\textemdash & \checkmark & $10.02 \pm 20.85$ & $1.42 \pm 1.10$ & $0.050 \pm 0.059$ \\
\checkmark & \checkmark & $6.41 \pm 11.73$   & $1.41 \pm 0.99$    
 & $0.050 \pm 0.059$ \\ \hline
\end{tabular}
    \label{tab:ablation_project}
\end{table}
Overall, these results confirm that the full architecture of TimeGazer—integrating all embedding components and both projection mechanisms—is essential for achieving stable and robust gaze prediction performance.

\section{Discussion}
This work represents the first attempt to explicitly exploit temporal modeling for gaze stabilization in AR systems. Through extensive experiments, we revealed several important aspects related to gaze refinement and task-driven visual attention.

\textbf{Temporal Modeling vs. Classical Gaze-Tracking Models:}\\
In contrast to classical approaches based on calibration correction or spatial filtering, temporal modeling methods exploit the implicit sequential dependencies embedded in gaze data. By leveraging historical trajectories, they can capture velocity, dispersion, and higher-order temporal correlations that guide the generation of more accurate and stable spatial positions. Both quantitative analyses from theoretical experiments (\cref{tab:theoratical_exp}) and subjective user survey (\cref{tab:user_survey}) support that temporal modeling enhances spatial compactness of gaze points during fixation, while improving alignment with intended target locations.

We attribute this advantage to the ability of sequence-to-sequence models to encode long-range dependencies from the pre-fixation search phase and align them with the subsequent fixation phase, thereby effectively bridging past dynamics and future stability \cite{Temporal_Modeling_Matters}. The consistently low variance of the Average Deviation (AD)—on the order of $10^{-3}$—further demonstrates the robustness of this approach across different users. This robustness suggests that temporal modeling captures a generalizable structure of gaze behavior, consistent with Yarbus’s classic assertion that “the pattern of eye movements is determined by the task” \cite{yarbus1967eye}. In particular, although our dataset consists of gaze sequences collected from 54 different participants, the seq2seq model successfully distilled shared temporal patterns that generalized well across individuals, demonstrating its applicability beyond subject-specific idiosyncrasies.

\textbf{Limitations:}\\
TimeGazer is built upon a sequence-to-sequence (seq2seq) temporal modeling framework that assumes a fixed-length input history, consistent model architecture, and predefined dataset conditions. In real-world scenarios, however, gaze sequences have variable lengths. To accommodate the model, historical sequences must be truncated, padded, or interpolated, which may reduce the precision of predicted fixation points and limit the model's ability to jointly capture the dynamics from target search to fixation phases. Consequently, the current model may not generalize directly to tasks such as reading or mid-air typing, which exhibit different temporal patterns and variable sequence lengths, and would require task-specific retraining using appropriately tailored historical sequences.

Additionally, in our data collection, some fixations start relatively close to the target, producing less stable gaze trajectories. These cases are more likely to be removed during data cleaning, which results in an underrepresentation of short-distance targets in the dataset. As a result, the model exhibits differential responsiveness to targets at varying distances, performing slightly better for medium-to-long-range fixations than for very short-range fixations.

\textbf{Futrue work:}\\
To address the limitation of fixed-length inputs, we plan to investigate adaptive sequence modeling approaches, such as hierarchical seq2seq architectures or attention-based transformers, which can process variable-length historical sequences without truncation or interpolation, thereby preserving the full temporal dynamics from target search to fixation. To mitigate the imbalance between near- and far-distance fixations, we aim to augment the dataset with additional short-distance target scenarios and incorporate a distance-aware weighting scheme during training. This strategy is expected to enhance model responsiveness and prediction accuracy across all fixation distances. Furthermore, we will explore integrating uncertainty estimation or robust regression techniques to better handle unstable or noisy fixation trajectories, thereby improving the reliability of predicted gaze points under diverse real-world conditions.

\section{Conclusion}

In this paper, our approach positions temporal sequence modeling as a foundational paradigm for gaze-based interaction in AR. By treating gaze stabilization not as a post-processing filter but as a predictive, task-driven temporal inference problem, we establish a new perspective on how the dynamics of human eye movements can be harnessed for neural modeling and facilitate interaction design.

Our TimeGazer model exemplifies this shift: it shows that latent temporal patterns in gaze trajectories carry rich cues about user intent and attentional stability, and that exploiting these cues can unlock performance gains unattainable through static or geometry-only approaches. Beyond the specific improvements in dispersion and accuracy, our approach reframes gaze interaction as a temporally grounded process, suggesting opportunities for adaptive interfaces, multimodal fusion strategies, and personalized attention-aware AR systems.

Looking ahead, temporal modeling of gaze could become a unifying principle for bridging perception and action in immersive environments—supporting not only robust object selection but also advanced tasks such as gaze-guided navigation, collaborative AR/VR workflows, and cognitive state estimation. By revealing the untapped potential of time-series dynamics, our study lays conceptual and practical groundwork for the next generation of gaze-based AR technologies.

%We presented TimeGazer, a temporal sequence modeling framework for gaze stabilization in AR. By predicting fixation-phase gaze points from historical trajectories, TimeGazer reduces spatial dispersion and improves alignment with targets. Experiments and user studies show clear advantages over classical gaze-tracking methods, with enhanced stability, accuracy, and interaction efficiency. Our work demonstrates the potential of temporal modeling for real-time, task-oriented gaze interaction in AR, paving the way for broader applications in immersive environments.

% The authors wish to thank A, B, and C. This work was supported in part by
% a grant from XYZ.}

%\bibliographystyle{abbrv}
\bibliographystyle{abbrv-doi}

\bibliography{GazeStabilization}

\appendix

% === 附录A ===
\setcounter{figure}{0}
\setcounter{equation}{0}
\setcounter{table}{0}
\renewcommand{\thefigure}{A.\arabic{figure}}
\renewcommand{\theequation}{A. \arabic{equation}}
\renewcommand{\thetable}{A. \arabic{table}}

\section{Analysis and Processing of Raw Gaze Data} 
\subsection{Calculate Visual Angle}\label{app:calculate_visual_angle}
We compute the visual angle $A$ of the AR device according to \cref{eq:visual_angle}:
\begin{equation}
A = 2 \arctan{\left(\frac{H}{2D}\right)},
\label{eq:visual_angle}
\end{equation}
where $D$ denotes the distance between the eye and the screen, and 
$H$ represents the diameter of the viewed object, assuming it can be approximated as a sphere.

\subsection{Define Angular Velocity Threshold}\label{app:calculate_angular_velocity}
\begin{figure}[!ht]
    \centering
    \includegraphics[width=0.8\columnwidth]{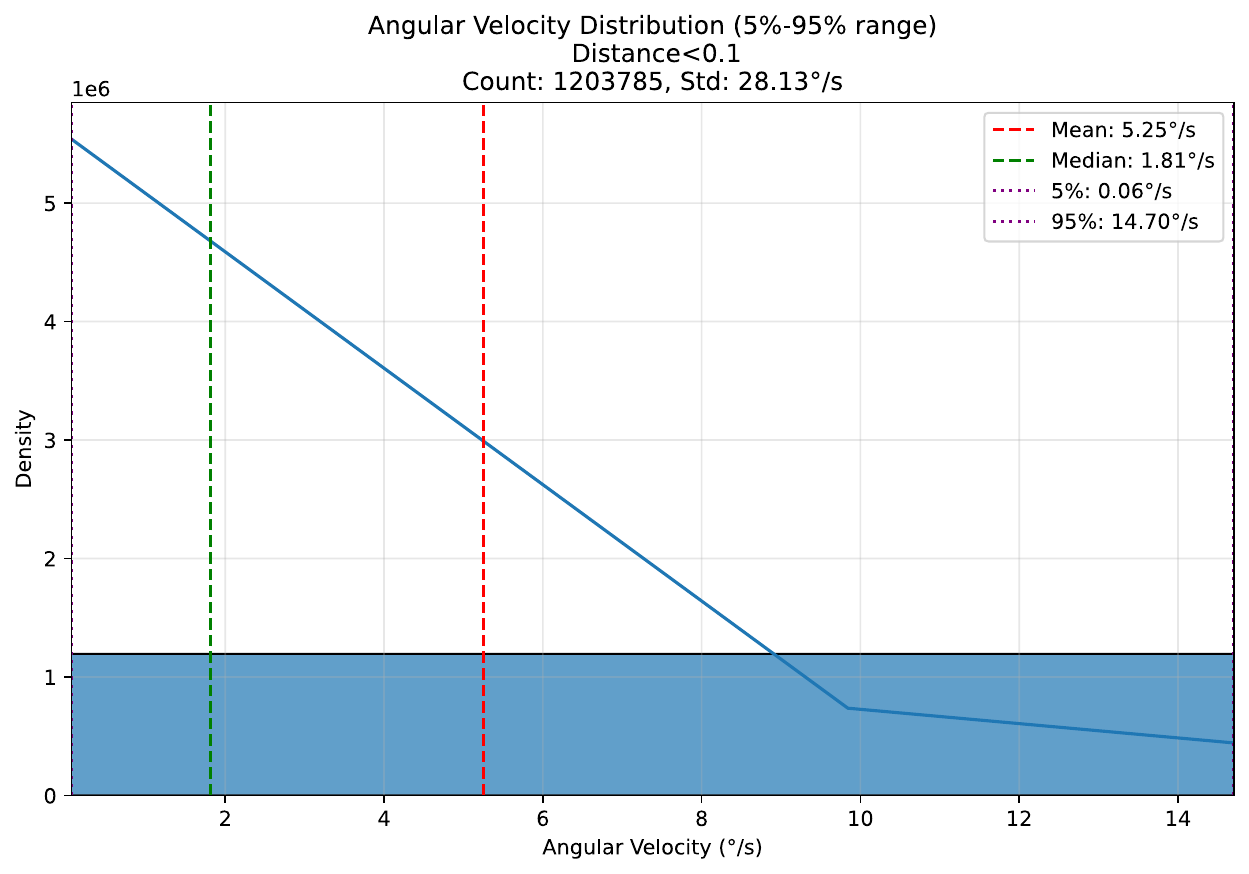}
    \caption{Angular velocity Distribution within effective fixation region.}
    \label{fig:app:angular_velocity}
\end{figure}

To define the angular velocity threshold that better suits our active gaze fixation task, we calculated the angular velocity distribution of gaze samples(range from 5\% to 95\%) within the effective fixation region defined in \ref{analysis_raw_data}. As shown in \cref{fig:app:angular_velocity}, the angular velocity of 95\% gaze points fell below 14.70°/s. 

\subsection{Define the Length of Historical Sequence}
\label{app:define_historical_sequence_length}
\begin{figure}[!ht]
    \centering
    \includegraphics[width=0.8\columnwidth]{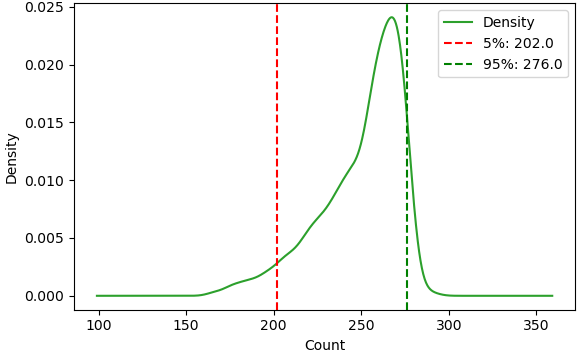}
    \caption{Density distribution of fixation-phase points, guiding the selection of a 96-point historical sequence.}
    \label{fig:app:fixation_phase_range}
\end{figure}

To set the historical sequence length for our seq2seq model, we analyzed fixation-phase distributions in the cleaned dataset. As shown in \cref{fig:app:fixation_phase_range}, over 95\% of sequences exceed 202 points, with total lengths ranging from 298 to 300. We therefore chose 96 points as a balanced length, ensuring representativeness while maintaining compatibility with other model modules.

\subsection{Parameter Sensitivity Analysis}\label{app:parameter_sensitivity_analysis}
To ensure the robustness of our parameter choices, we conducted a sensitivity analysis by training and testing the proposed model (\cref{sec:model}) on datasets generated under different configurations of angular velocity thresholds. The evaluation metrics are consistent with those defined in \cref{sec:metrics}. In the result table, bold numbers denote the best performance, while underlined numbers indicate the second-best performance.

We first fixed the fixation window size at 200 ms and varied the angular velocity threshold. As shown in \cref{tab:validation_angular_threshold}, the configuration using a threshold of 20°/s achieved the highest overall performance. Furthermore, this threshold encompasses more than 95\% of gaze samples within the defined fixation region, supporting its selection as the default angular velocity threshold.
\begin{table}[!h]
\centering
    \caption{Validation on angular velocity threshold. Values are reported as mean $\pm$ standard deviation (SD).}
\begin{tabular}{cccc}
\hline
Velocity & CI$\uparrow$& AI$\uparrow$& AD$\downarrow$\\ \hline
5°/s      & \textbf{3.61}±5.35& 1.31±0.91& \textbf{0.043}±0.033\\
10°/s     & 3.28±8.35& 1.37±1.09& 0.048±0.016\\
15°/s     & 2.97±4.04& 1.40±1.03& 0.048±0.055\\
20°/s     & \underline{3.44}±5.35& \underline{1.44}±1.05& \underline{0.048}±0.055\\
25°/s     & 2.91±6.72& \textbf{1.49}±1.37& 0.049±0.053\\
30°/s     & 2.87±3.95& 1.32±0.91& 0.049±0.045\\ \hline
\end{tabular}
\label{tab:validation_angular_threshold}
\end{table}

\subsection{Synthetic Sequence Generation}
\label{app:synthetic_data_and_blending}
Given a real fixation subsequence $S^r = \{(x_t,y_t,v_t)\}_{t=1}^{T}$ from one trial and the corresponding target position $\mathbf{g} = (x_g,y_g)$, we first determine the boundary index $T_B$ that separates the target-searching phase from the target-fixation phase, based on the angular velocity threshold and fixation window size.

To emulate an idealized fixation behavior, we construct a synthetic sequence
\begin{equation}
S^s = \{(x_t,y_t,v_t)\}_{t=1}^{T_B} \;\Vert\; \{(x'_t,y'_t,v'_t)\}_{t=T_B+1}^{T},
\end{equation}
where $\Vert$ denotes sequence concatenation, and $(x'_t,y'_t,v'_t)$ are generated by applying spatial contraction and translation toward the target for points in the fixation phase:
\begin{equation}
    \left\{
             \begin{array}{lr}
                 (x_t^{\prime},y_t^{\prime})=\mathbf{g} + \beta \cdot (x_t,y_t), & \beta \in {(0,1)}. \\
                  v_t^{\prime} = \frac{(x_{t}^{\prime},y_{t}^{\prime})-(x_{t-1}^{\prime},y_{t-1}^{\prime})}{\Delta t}, & \quad t = T_B+1,\dots,T. 
             \end{array}
    \right.
\end{equation}
Here, $\beta$ controls the contraction strength (smaller $\alpha$ indicates stronger convergence toward the target), and $\Delta t$ denotes the sampling interval at 60Hz.

This procedure yields synthetic fixation subsequences with reduced spatial deviations and stronger internal cohesion, reflecting the idealized convergence pattern expected during the target-fixation phase.

\end{document}